\documentclass[12pt]{article}
\usepackage{graphpap,epsfig,amssymb,graphicx,textcomp}
\usepackage[utf8]{inputenc}
\usepackage[english]{babel}
\begin {document}

\begin {center}
{\bf {\Large Universality class of the nonequilibrium phase transition in two dimensional Ising ferromagnet driven by propagating magnetic field wave}}
\end {center}
\begin {center}{ Ajay~Halder$^\star$ and Muktish~Acharyya$^\dagger$}
\\{Department of Physics, Presidency University}\\
{86/1 College Street, Kolkata-73, India}
\\{$^\star$ ajay.rs@presiuniv.ac.in}
\\{$^\dagger$ muktish.physics@presiuniv.ac.in}
\end {center}
\vskip 0.2cm

\noindent {\bf Abstract:} The finite size analysis of the nonequilibrium 
phase transition, in two dimensional Ising 
ferromagnet driven by plane propagating magnetic
 wave,
 is studied by Monte Carlo simulation. It is observed that the system undergoes a nonequilibrium dynamic
 phase transition from a high temperature dynamically symmetric ({\it propagating}) phase to a low temperature dynamically symmetry-broken ({\it pinned}) phase as the system is cooled below 
the transition temperature. This transition temperature is 
determined precisely by studying the
fourth order Binder Cumulant of the dynamic order parameter as a function
of temperature for 
different system sizes ($L$). From the finite size analysis of
dynamic order parameter ($Q_L \sim L^{-{{\beta} \over {\nu}}}$) and 
the dynamic susceptibility ($\chi^{Q}_{L} \sim L^{{\gamma} \over {\nu}}$), we have estimated the critical exponents
$\beta/\nu=0.146\pm0.025$ and $\gamma/\nu=1.869\pm0.135$
(measured from the data read at the critical temperature obtained
from Binder cumulant), and $\gamma/\nu=1.746\pm0.017$ (measured from
the peak positions of dynamic susceptibility). Our results
indicate
 that such driven Ising ferromagnet belongs to the same universality class
 of the two-dimensional equilibrium Ising ferromagnet
(where $\beta/\nu=1/8$ and $\gamma/\nu=7/4$), 
within the limits of statistical errors.

\vskip 0.6cm


\newpage

{\noindent \bf 1. Introduction}

The driven Ising ferromagnet shows interesting nonequilibrium phase transitions
\cite{revrmp,revijmpc}. This time dependent drive may be of two kinds:
(i) an applied magnetic field which is oscillating in time and uniform over
the space at any particular instant (ii) the applied magnetic field has
a spatio-temporal variation which may be the type 
of propagating or standing magnetic
field wave. The first kind of driving magnetic field has drawn much attention
to the researchers and a considerable volume of studies was done in this
direction, in last two decades. Here, a few of those may be mentioned as 
follows: (i) the critical slowing down and the divergence of the specific
heat near the dynamic transition temperature\cite{ma1}, (ii) the divergence
of the fluctuations of the dynamic order parameter\cite{ma2}, (iii) the
growth of critical correlation near the dynamic transition temperature
\cite{rikvold1}. These studies are an integrated effort to establish that the 
nonequilibrium transition in kinetic Ising ferromagnet driven by oscillating
magnetic field is indeed a thermodynamic {\it phase transition}. 

The nonequilibrium phase transitions in other magnetic models
(e.g., Blume-Capel, Blume-Emery-Griffiths models etc.) driven by
oscillating (in time but uniform over the space) magnetic field 
have been studied \cite{oth1,oth2,erol1} also in last few years to 
present some interesting nonequilibrium behaviours.
The nonequilibrium phase transitions were studied in
\cite{mx1,mx2,mx3,mx4,mx5} mixed spin systems
driven by oscillating magnetic field, recently.

The another kind of external drive may be the magnetic field with
spatio-temporal variation. The prototypes of these  spatio-temporal
drives are propagating or standing magnetic field waves. 
In the last few years, a number of investigations, on the nonequilibrium
phase transitions in Ising ferromagnet driven by propagating
and standing magnetic field wave, are done\cite{ma3,ma4,ah1,ah2,erol2}
through Monte Carlo methods. Here, the essential findings are the 
nonequilibrium phase transitions between two phases, namely, the low
temperature ordered 
pinned phase (where the spins do not flip) and a high temperature
disordered phase where a coherent propagation
(in the case of propagating magnetic field wave) or coherent 
oscillation (in the case
of standing magnetic field wave) of spin bands are observed.
The transitions are marked by the divergences of dynamic susceptibility
near the transition point. 

However, the detailed finite size analysis, were not yet performed to know
the {\it universality class} of this nonequilibrium phase transition observed in
Ising ferromagnet driven by propagating magnetic field wave. This is the
key issue of the present study.

In this paper, we have investigated the nonequilibrium behaviour and the finite size effect
 of spin-$\frac{1}{2}$ Ising ferromagnet 
under the influence of propagating magnetic wave by Monte Carlo methods.
 The paper is organised as follows: 
The model and the MC simulation technique are discussed in Sec. II, the numerical 
results are reported in Sec. III and the paper ends with a summary in Sec. IV.

\vskip 0.6cm

\noindent {\bf 2. Model and Simulation}
\vskip 0.5cm

The {\it time dependent} Hamiltonian of a two dimensional driven Ising ferromagnet 
is represented by, 
\begin{equation}
 H(t) = -J\Sigma\Sigma' s^z(x,y,t) s^z(x',y',t) - \Sigma h^z(x,y,t)s^z(x,y,t) .
\end{equation}
Here $s^z(x,y,t)=\pm 1,$ is the Ising {\it spin variable} at lattice site $(x,y)$ at time $t$. 
The summation  $\Sigma'$ extends over the nearest neighbour sites $(x',y')$ of a given site $(x,y)$.
 $J(>0)$ is the {\it ferromagnetic spin-spin interaction strength} between the nearest neighbour pairs of Ising spins. 
For simplicity, we have considered the value of $J$ to be uniform over the whole lattice.
 The externally applied driving {\it magnetic field}, is denoted by $h^z(x,y,t)$, at site $(x,y)$ at time $t$.
$h^z(x,y,t)$ has the following form for propagating magnetic wave, 
\begin{equation}
 h^z(x,y,t)= h_0 cos \{2\pi (ft-\frac{x}{\lambda})\} 
\end{equation}

Here $h_0$ and $f$ represent \textit{the field amplitude and the frequency}
respectively
 of the propagating magnetic wave,
 whereas $\lambda$ represents \textit{the wavelength} of the wave.
The wave propagates along the $X$-direction through the lattice.

An $L\times L$ square lattice of Ising spins is taken here as a model system. The boundary conditions 
applied at both directions are {\textsl{periodic}} which preserve the translational invariances in the system.
 Using {\it Monte Carlo Metropolis single spin flip algorithm} with 
{\it parallel} updating rule\cite{springer}, the dynamics of the
system are simulated.
 The initial state of the system 
is chosen as the high temperature random disordered phase,
 in which, at any lattice site, both the two states $(\pm 1)$ of the 
Ising spins have equal probabilities.
The system is then slowly cooled down to any lower temperature $T$ and the dynamical quantities are calculated.
 The {\it Metropolis probability}\cite{springer} of single 
spin flip at temperature $T$ is given by,
\begin{equation} 
W((s^z)_i\to (s^z)_f) = {\rm Min} [{\rm exp}(\frac{-\Delta E}{k_BT}),1] 
\end{equation}
 where $\Delta E$ is the energy change due to spin flip from $i$-th state to $f$-th state and 
$k_B$ is the Boltzman constant. Parallel updating of $L^2$ spin states in an 
$L\times L$ square lattice constitute the unit time step and is called 
{\it Monte Carlo Step per Spin} (MCSS). The applied magnetic field and 
the temperature are measured in the units of $J$ and $J/k_B$ respectively. 

\vskip 1cm

\noindent {\bf 3. Results}
\vskip 0.2cm

The nonequilibrium behaviour of the two dimensional
 Ising ferromagnet is studied here
 in $L\times L$ square lattices of different sizes ($L$)
 where a propagating magnetic wave is passing through the system. 
The frequency $(f)$ of magnetic field oscillation,
 wavelength $(\lambda)$ of the magnetic wave and the amplitude $(h_0)$ of field strength
 are kept constant throughout the present
study. These constant values are respectively
 $f=0.01~(MCSS)^{-1}$, $\lambda=16~$ lattice units and $h_0=0.3~J$. 
For $f=0.01$,  100 MCSS is required for a complete time cycle.
 The finite size effect is studied by taking into account 
four different system sizes (within the limited computational facilities
available) such as
 $L=16,~32,~64~and~128$. The system 
(for any fixed value of $L$)
has been cooled down in small steps 
$(\Delta T=0.005~J/{k_B})$ from high temperature phase, i.e. the dynamical disordered phase, to 
reach any dynamical steady state at temperature $T$.
 The dynamical quantities are calculated when the system has achieved 
the nonequilibrium steady state.
 For this we have kept the system in constant temperature for a sufficiently long time; 
$12000$ (for $L=128$) to $32000$ (for $L=16$) cycles of magnetic oscillations and discarding the
 initial (or transient) $1000$ cycles and taking average over the remaining cycles. 
We have detected a dynamical phase transition from high temperature symmetric propagating (spin bands) phase to
 low temperature symmetry-broken pinned phase.
 The dynamic {\it Order parameter} for the phase transition 
is defined as the average magnetisation 
per spin over a full cycle of external magnetic field, i.e.
\begin {equation}
 Q=f\times\oint M(t)dt,
\end{equation}
 where $M(t)$ is the value of instantaneous magnetisation per 
spin at time $t$ which can be obtained as
\begin{equation}
 M(t)=\frac{1}{L^2}\sum_{i=1}^{L^2} s^z(x,y,t).
\end{equation}
At very high temperature, the flipping probability of the spin, is quite high
 alongwith the oscillation of the magnetic field. As 
a result the value of the instantaneous magnetisation is almost close to zero.
Consequently, by definition, 
the value of the dynamic order parameter is very small,
 thus identifying the dynamically disordered propagating phase $(Q=0)$
(see fig-1b). 
It may be noted here, that the instantaneous magnetisation fluctuates 
symmetrically about zero (see fig-1d). Hence, this may be
characterised as a dynamically symmetric phase.
As the system is cooled down below the critical temperature, which depends on the value of 
magnetic field strength, the flipping probability of the spin gets 
reduced; also the magnetic field strength may not be adequate to flip the spins
 and the spins are locked or pinned in a particular orientation 
giving rise to a large and nearly steady value of 
average magnetisation. 
This phase is identified as the dynamically ordered or 
pinned phase $(Q\ne 0)$ (see fig-1a).
Unlike, the dynamically symmetric phase (mentioned above), here the 
instantaneous magnetisation varies asymmetrically about zero
(see fig-1c). So, this may be
called a dynamically symmetry broken phase.
 The variation of the order parameter for the dynamic phase transition (DPT) for
 four different system sizes are shown in the fig.2a.

The dynamical critical point is determined with high precision by studying
 the thermal 
variation of fourth order Binder cumulant 
($U_L(T) =1-\frac{\langle Q^4 \rangle_L}{3\langle Q^2\rangle_L^2}$)
of dynamic order parameter $Q$ 
for different system sizes ($L$).  
 Fig.2b shows the variation of the Binder Cumulant 
($U_L(T)$) with temperature ($T$) 
for different values of $L$. From this figure we 
have determined the value of critical temperature 
as $T_d=2.011~J/{k_B}$, which is the value
 of temperature where the Binder cumulants for 
different lattice sizes have a common intersection. 
 Now it is known from the behaviour of the kinetic Ising model 
that the scaled variance of the dynamical order parameter may be regarded as the susceptibility of 
the system, which can be defined as follows:
\begin{equation}
{\chi}_L^Q=L^2(\langle Q^2\rangle-{\langle Q \rangle}^2).
\end{equation}
 Fig.2c. shows the variation of the scaled variance with the temperature.
 As we see from the figure that the susceptibility gets peaked near the dynamical transition temperature 
showing the tendency of divergence near $T_d$, as the system size increases.
Now we adopt the {\it finite-size scaling} analysis 
to determine the critical exponents for the two dimensional kinetic Ising 
ferromagnet driven by magnetic wave. For this reason we use the usual 
technique of expressing the measured 
quantities as a function of the system size. We assume  
the following scaling forms for the order parameter $Q$ and susceptibility $\chi^Q$ at the critical 
 temperature:
\begin{equation}
{\langle Q \rangle}_L\propto L^{-\beta/\nu}
\end{equation}
\begin{equation}
{\chi}^Q_L\propto L^{\gamma/\nu}.
\end{equation}

 It has to be noted here that though we do not have any value measured at
 the critical 
temperature which has been determined (as common intersection) 
from the Binder cumulant versus temperature curves for 
different $L$, the values of $Q$ \& $\chi^Q$ 
have been read out from the respective graphs which
 represent the average values at any temperature. 
Moreover, the detailed investigations done previously 
\cite{erol1}, show that the above scaling forms are
 also applicable to classify the universality 
classes of the driven magnetic systems. 
Fig.3a. shows the log-log plot of the dynamic order parameter
$\langle Q\rangle_L$ as a function of the linear system size $L$ at the dynamic transition 
temperature. The value of the critical exponent, as 
estimated from this simulational study, 
is $\beta/\nu=0.146\pm 0.025$ for the dynamic order parameter.
From the log-log plot fig.3b. of the susceptibility ${\chi}^Q_L$ or 
the scaled variance of the order parameter
 $\chi^Q_L$ as a function of linear system size $L$
 we obtained the estimate of the value of the 
critical exponent $\gamma/\nu$. The 
values are $\gamma/\nu=1.869\pm 0.135$ 
(using the data obtained at $T_d=2.011~J/k_B$ from the respective graphs)
 and $\gamma/\nu=1.746\pm 0.017$ (using the data obtained at 
the peak position of susceptibility).
 It is interesting to note that these 
estimated values of the critical exponents, for the 
two dimensional driven Ising ferromagnet,  are very close to those 
of the two dimensional equilibrium 
Ising ferromagnet, which are $\beta/\nu=1/8=0.125$ and 
$\gamma/\nu=7/4=1.75$ \cite{Huang}.

\newpage
\vskip 0.6cm

\noindent{\bf 4. Summary:}
\vskip 0.2cm
In this study we have mainly focused our attention on the finite size analysis and 
the critical aspects of the dynamic phase transition near the dynamic transition temperature
of an $L\times L$ square type Ising ferromagnet driven by propagating magnetic wave.
 We have taken four different sizes of square lattice ($L=16,~32,~64~and~128$).
 We have simulated the results using Monte Carlo methods using the Metropolis
single spin flip algorithm with parallel updating rules. 
Our findings suggest that, within the limits of statistical errors 
obtained in this study, the 
estimated values of the critical exponents 
near the dynamic transition temperature
are very close to those for the two dimensional
 equilibrium Ising ferromagnet. 
As concluding remarks, 
we state that the nonequilibrium phase transition, observed in 
the two dimensional Ising ferromagnet driven by
 magnetic field wave, belongs to the same universality class of 
equilibrium two dimensional Ising 
equilibrium ferro-para phase transition.

\vskip 1cm

\noindent {\bf Acknowledgements:} MA thanks Chandan Dasgupta for helpful
discussion and acknowledges financial support through FRPDF grant provided 
by Presidency University.

\vskip 1cm

\noindent{\bf V. References:}

\vskip 0.6cm
\footnotesize{
}
\newpage

\begin{center}{\bf Figure captions}\end{center}

\vskip 0.5cm

\noindent Figure-1. 
Fig.1a. \& fig.1b. show the lattice morphology of
the pinned and propagating phase respectively at time $t=39937~MCSS$ for $L=64$.
 Fig.1c. and fig.1d. show the dynamical symmetry breaking (change in the value of 
average magnetisation per spin from non-zero to nearly zero value). Fig.1a. \& fig.1c.
 are at temperature $T=1.8$ whereas fig.1b. \& fig.1d. are at temperature $T=2.5$

\vskip 0.5cm

\noindent Figure-2.  Temperature variation of different quantities for different values of linear system size $L$: (a) Order parameter $Q$, (b) Binder cumulant $U_L$
, (c) scaled variance of order parameter $varQ$ or susceptibility ${\chi}^Q_L$.

\vskip 0.5cm

\noindent Figure-3.  Log-log plot of (a) order parameter $Q$ and 
(b) scaled variance $varQ$ or susceptibility ${\chi}^Q_L$ as a 
 function of linear system size $L$. In (b) red dots represent the value of susceptibility 
at $T_d$ whereas blue triangles represent the same at peak positions.

\newpage

\begin{figure}[h]
\begin{center}
\begin{tabular}{c}
\resizebox{7cm}{7cm}{\includegraphics[angle=0]{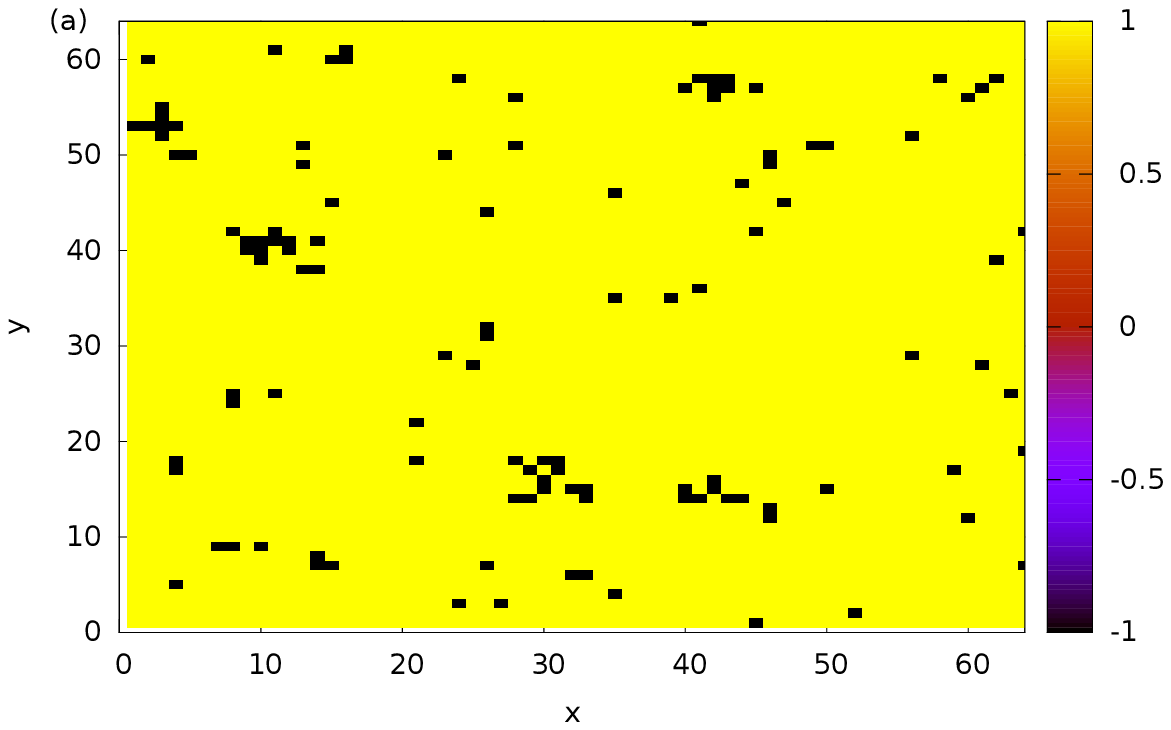}}
\resizebox{7cm}{7cm}{\includegraphics[angle=0]{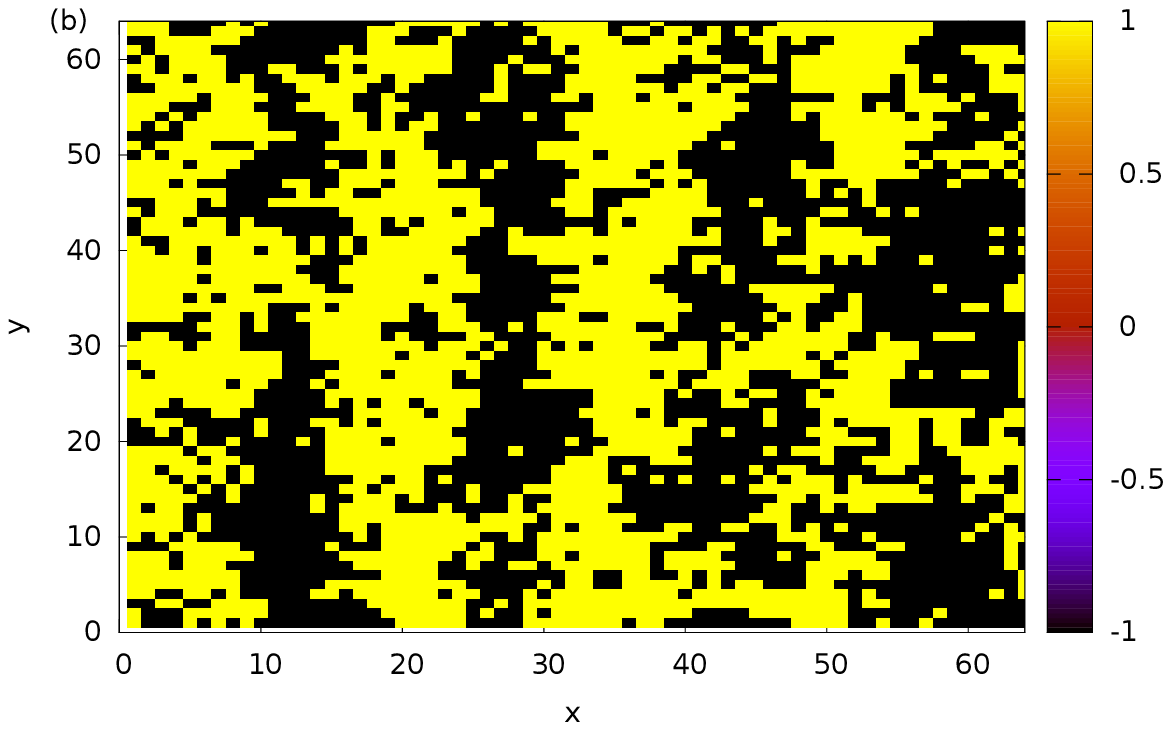}}
\\
\resizebox{7cm}{7cm}{\includegraphics[angle=0]{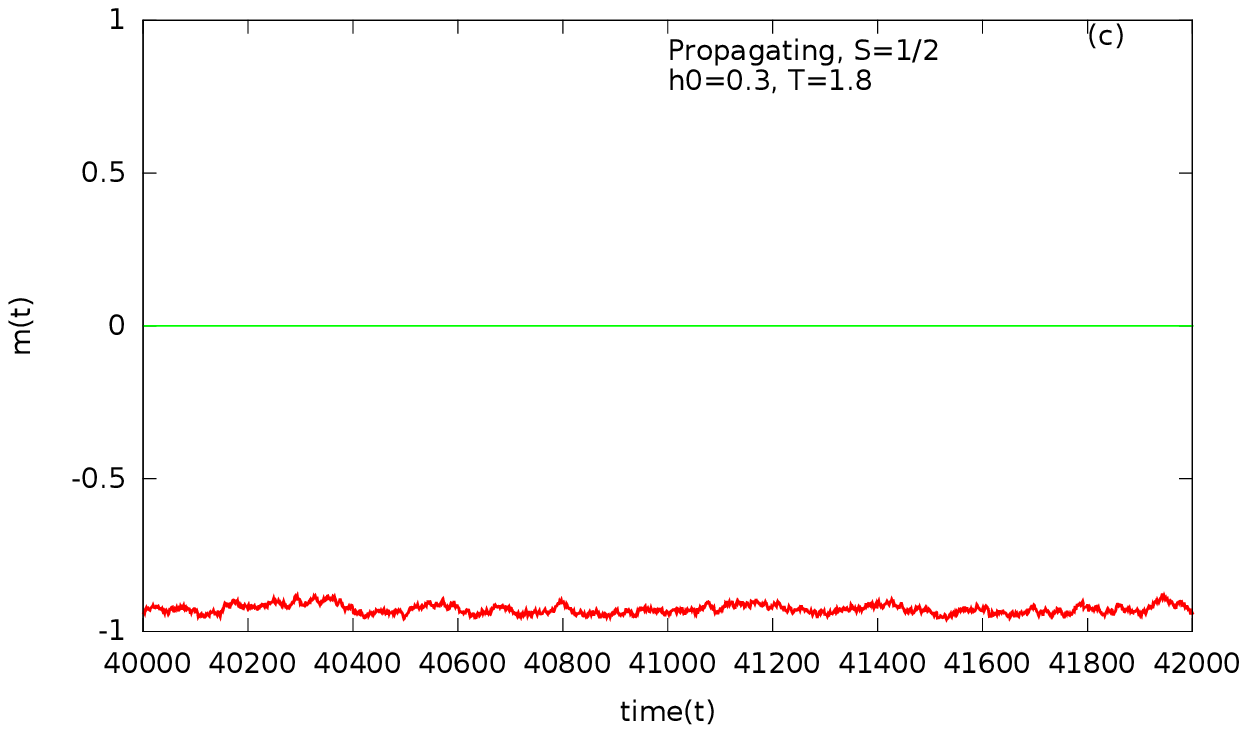}}
\resizebox{7cm}{7cm}{\includegraphics[angle=0]{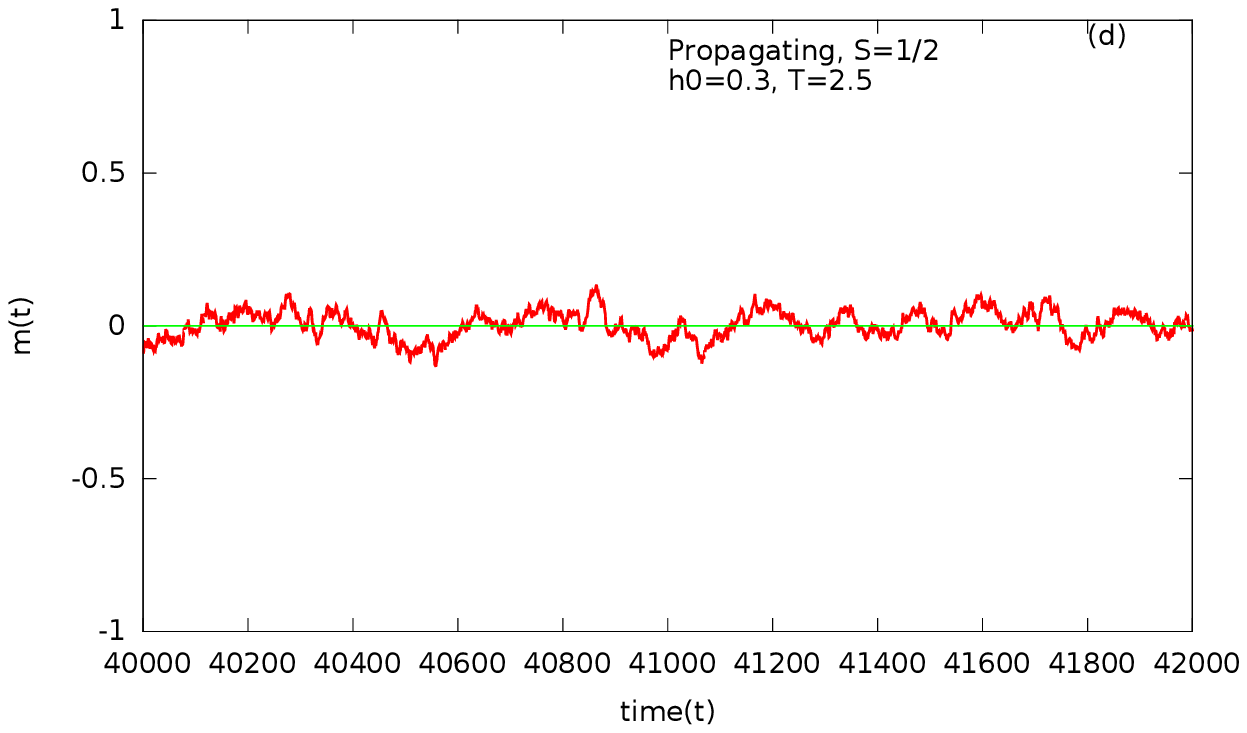}}
          \end{tabular}

\begin{center}{Figure-1}\end{center}
\end{center}
\end{figure}
\newpage

\begin{figure}[h]
\begin{center}
\begin{tabular}{c}
{\includegraphics[angle=0]{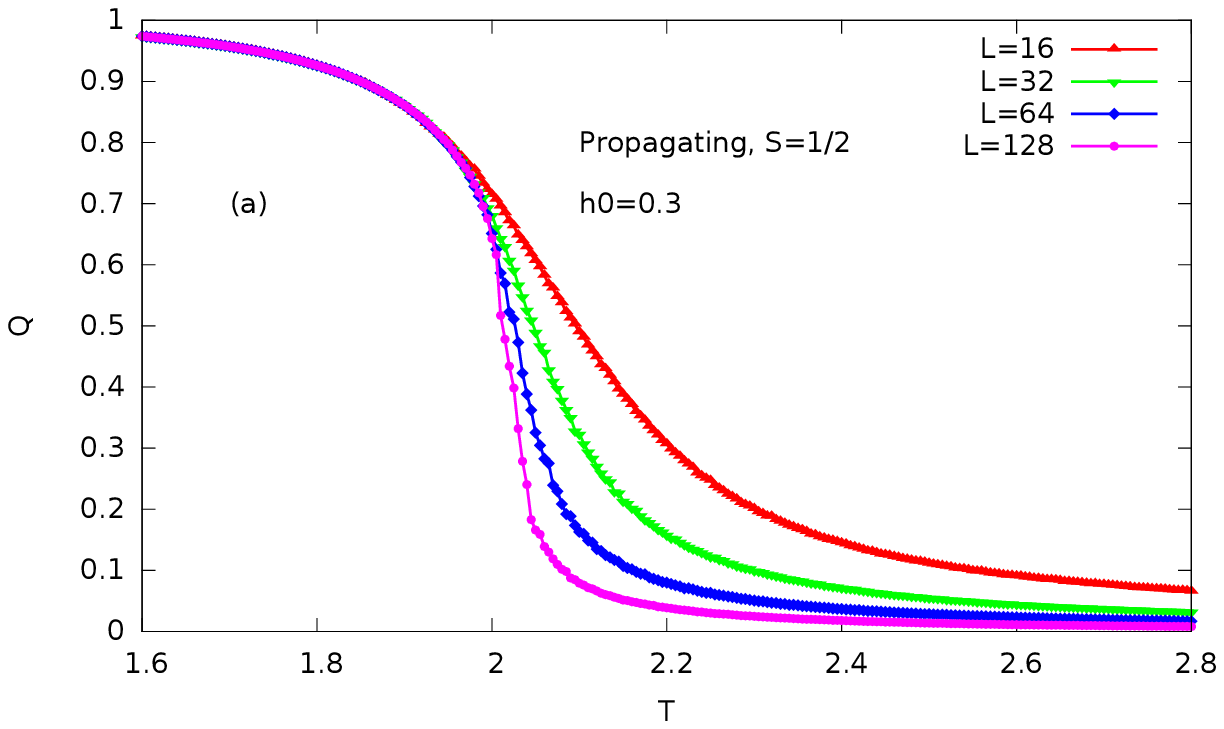}}\\
{\includegraphics[angle=0]{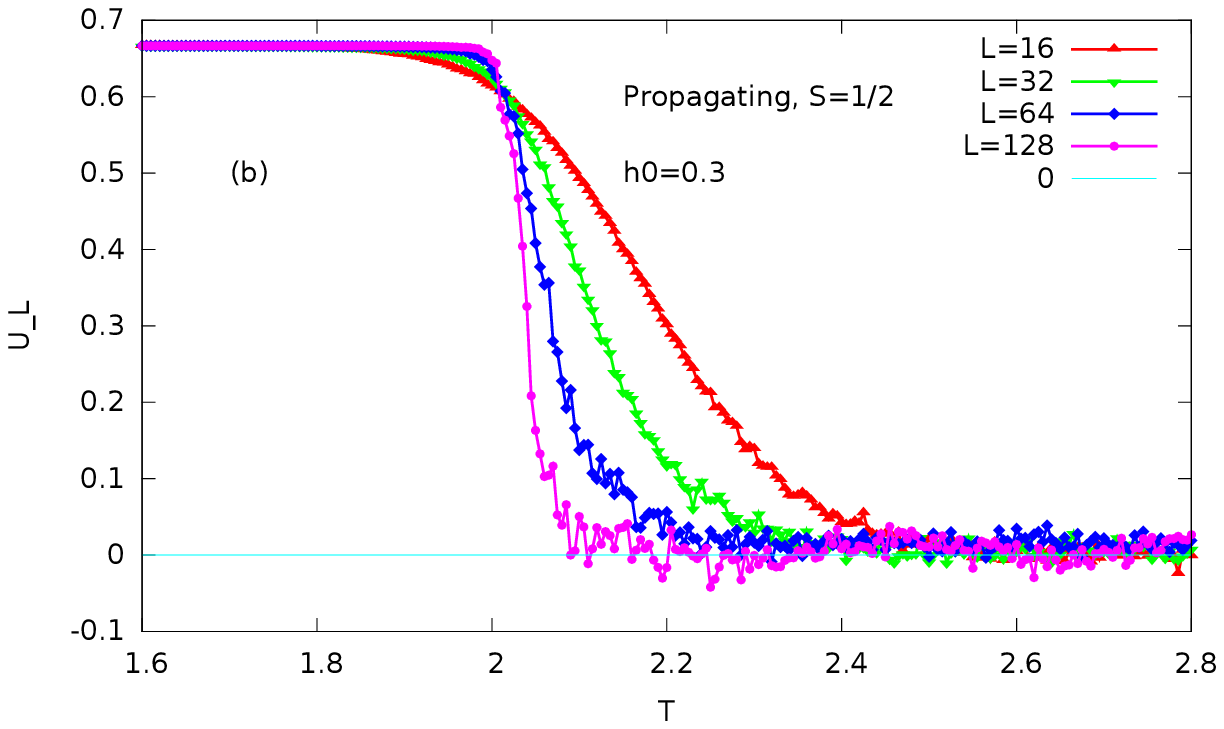}}
\\
{\includegraphics[angle=0]{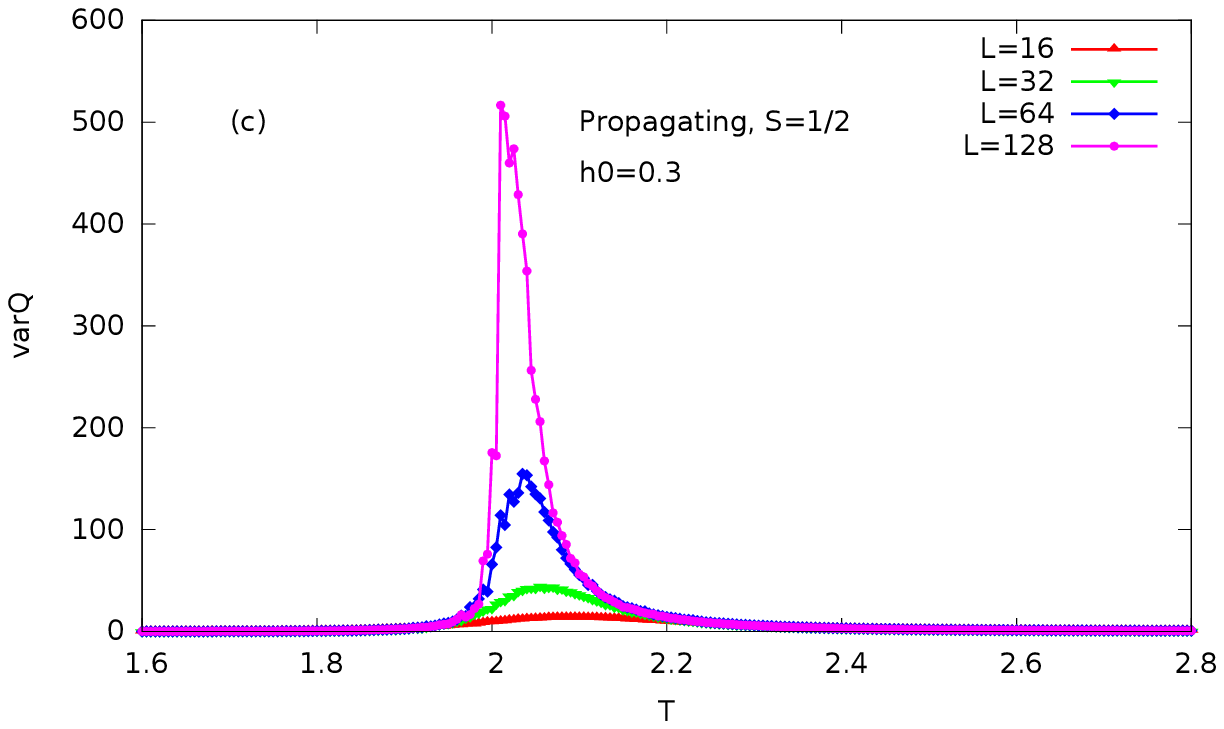}}
          \end{tabular}

\begin{center}{Figure-2}\end{center}
\end{center}
\end{figure}

\newpage

\begin{figure}[h]
\begin{center}
\begin{tabular}{c}
{\includegraphics[angle=0]{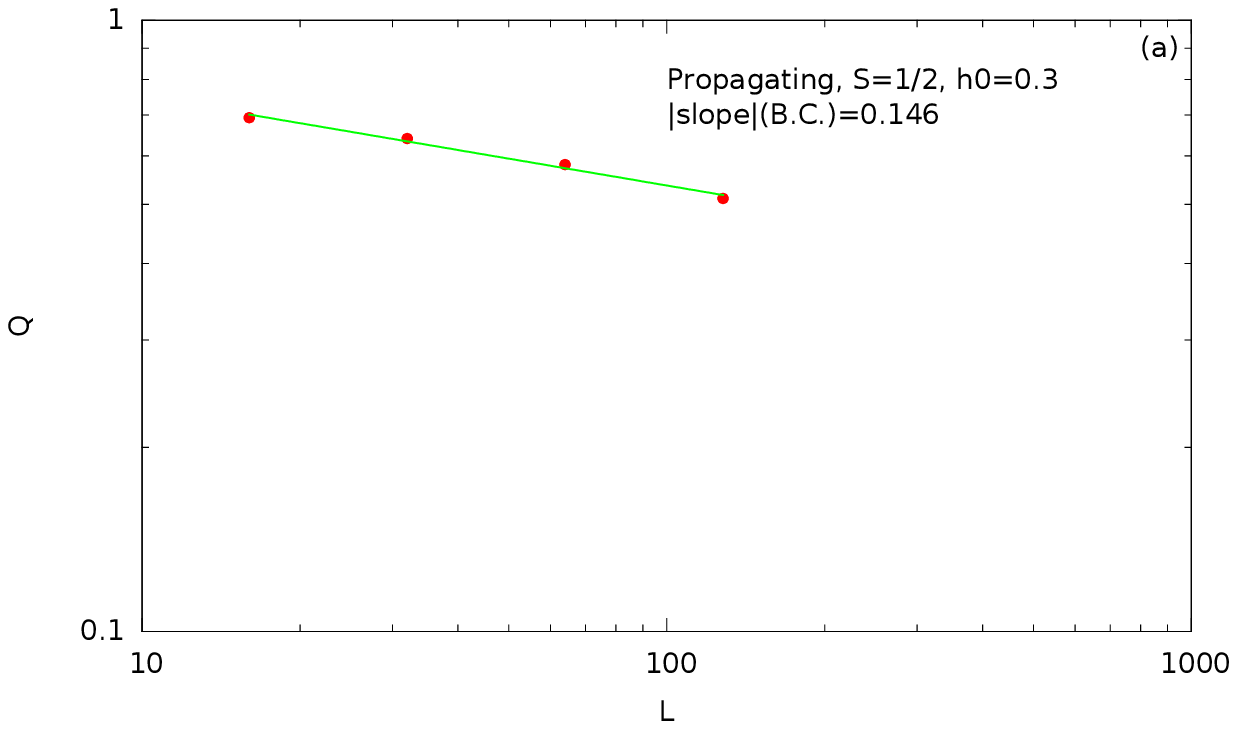}}\\
{\includegraphics[angle=0]{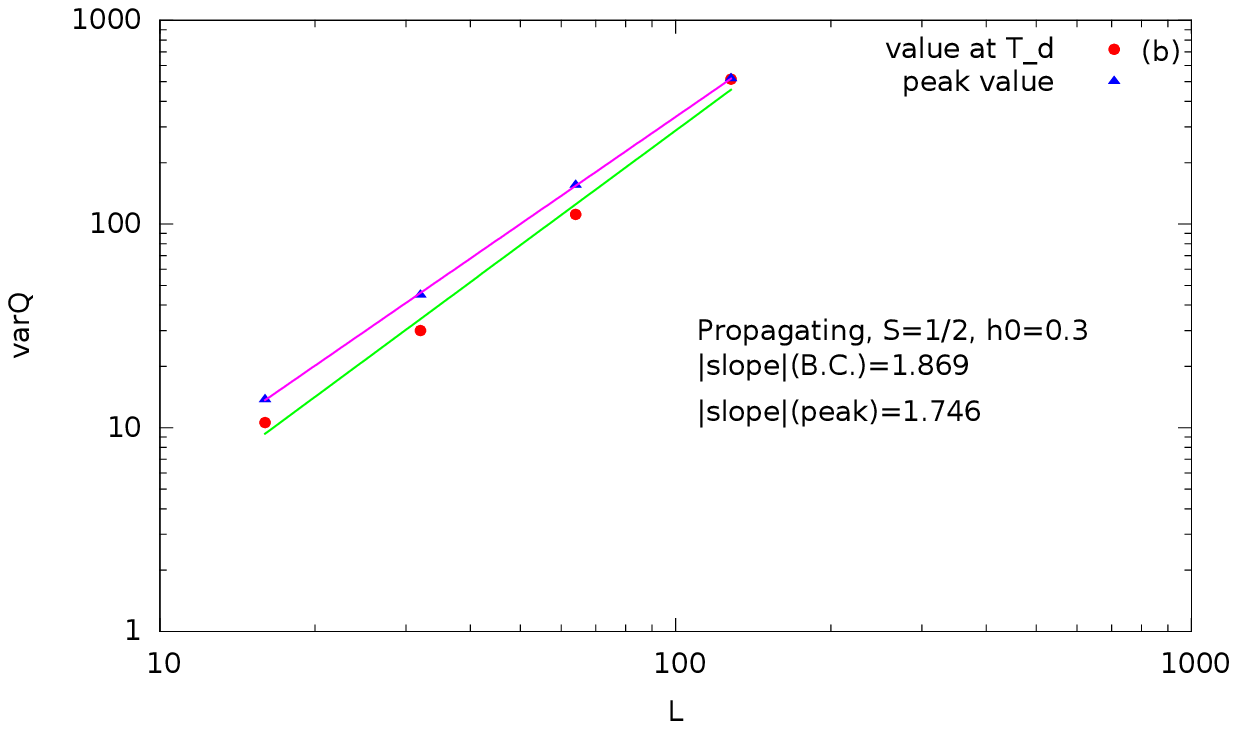}}
\\
          \end{tabular}
%

\begin{center}{Figure-3}\end{center}
\end{center}
\end{figure}


\end{document}